\documentclass{vgtc}                          

\graphicspath{{figures/}{pictures/}{images/}{./}} 

\usepackage{times}                     

\usepackage{tabu}                      
\usepackage{booktabs}                  
\usepackage{lipsum}                    
\usepackage{mwe}     
\usepackage{tabularx}

\usepackage{mathptmx}                  
\usepackage{xcolor}
\usepackage{soul}

\usepackage{pdfpages}

\usepackage{amssymb}
\usepackage{colortbl}

\usepackage{makecell}

\newcommand{\etal}{et al.}
\newcommand{\etals}{et al.'s}

\newcommand{\eg}{{e.g.,}}
\newcommand{\etc}{{etc.}}

\newcommand{\credit}{CRediT}
\newcommand{\creditfair}{CRediT-fAIR}

\newcommand{\authormara}{Mara}
\newcommand{\authorwesley}{Wesley}
\newcommand{\authorandrew}{Andrew}

\newcommand{\redacted}[1]{REDACTED}

\newcommand{\numParticipants}{21}

\newcommand{\secref}[1]{\hyperref[#1]{Sec.~\ref*{#1}}}
\newcommand{\appendixref}[1]{\hyperref[#1]{Appendix~\ref*{#1}}}
\newcommand{\figref}[1]{\hyperref[#1]{Fig.~\ref*{#1}}}
\newcommand{\eqnref}[1]{\hyperref[#1]{Eqn.~\ref*{#1}}}
\newcommand{\tabref}[1]{\hyperref[#1]{Table ~\ref*{#1}}}

\newcommand{\hlc}[2][yellow]{{%
      \colorlet{foo}{#1}%
      \sethlcolor{foo}\hl{#2}}%
}
\definecolor{quoteColor}{HTML}{002145}
\newcommand\qt[1]{\hlc[quoteColor!10]{``\textit{#1''}}}
\definecolor{wwQuoteColor}{HTML}{D6001C}
\newcommand\wwqt[1]{{\hlc[wwQuoteColor!10]{``\textit{#1''}}}}
\definecolor{exampleColor}{HTML}{9370DB}

\newcommand{\pxx}[1]
{\textbf{P}$_{\textrm{#1}}$}

\newcommand{\parahead}[1]
{%
  \paraheadd{#1}.
}

\newcommand{\paraheadd}[1]
{%
  \vspace{0.07in}%
  \noindent%
  \textbf{\textit{#1}}%
}

\def\subsubsec#1
{\subsubsection{#1}}

\onlineid{1020}

\vgtccategory{Research}

\vgtcinsertpkg

\title{Considering Contribution Statements in Visualization and HCI Research}

\usepackage{orcidlink}

\author{Mara Solen \orcidlink{0000-0002-5191-5193}\thanks{e-mail: marasolen@gmail.com}\\ %
        \scriptsize University of British Columbia %
\and Wesley Willett \orcidlink{0000-0002-6793-3062}\thanks{e-mail: wj@wjwillett.net}\\ %
     \scriptsize University of Calgary %
\and Andrew McNutt \orcidlink{0000-0001-8255-4258}\thanks{email: andrew.mcnutt@utah.edu}\\ %
     {\scriptsize University of Utah}}

\abstract{
Contribution statements are an increasingly common way to make research labor visible, reduce academic malfeasance, and provide broader transparency. Despite this potential value, they remain uncommon in visualization and HCI. To explore this gap, we conducted an online study with (N=\numParticipants) visualization and HCI researchers. We find a range of differing opinions about the utility of contribution statements, which are set against a background of tensions relating to contribution frameworks that inadequately fit contribution types in HCI and especially visualization, power dynamics between authors, bias in authorship perceptions, and the tedium of providing yet another form of documentation. From these factors, we offer a modest recommendation to authors: consider contribution statements. There are contexts when they may usefully explicate work, and others where they can cause author-team conflict or become a burdensome chore. Regardless of whether they are used by readers or not, we suggest that scaffolded mechanisms for reflecting on contribution roles are valuable both for public accountability and internal alignment. To institutions, we recommend that contribution statements be exempt from page or word limits, and that flexible templates and examples be provided to authors, but that they continue to not be required. By surfacing these perspectives, we seek to open a dialogue about what constitutes authorship, how our community might move toward more equitable and transparent attribution practices, and where the visualization and HCI communities might be uniquely equipped to help.
} 


\begin{document}


\firstsection{Introduction}

\maketitle

There are a range of differing standards for what counts as a contribution to a visualization paper.
Some papers include a multitude of different parties holding many differing and separate roles, while others may involve just one or two people who together hold all roles.
In parallel with these varied practices are varied sets of assumptions about who did what on a paper. It is easy, for instance, to see a prominent or famous name and assume that person did all the work, when in fact they may have merely attended a few meetings or provided a small amount of asynchronous feedback.

In some fields, of which medicine and bio-affiliated fields are common examples, these ambiguities are resolved through the notion of a contribution statement. These structures typically appear at the end of a paper and list a series of roles that each author performed in support of a paper (\figref{fig:statements}). For instance, \textit{Lisa Simpson: Ideation, Conceptualization, Writing; Homer Simpson: Supervision; Montgomery Burns: Funding.}
While these practices have not been embraced in visualization or human-computer interaction (HCI) aside from occasional use \cite{akbaba2026designing, bai2026shadow, cui2026codesigning, dhawka2024better, pearman2026threetopo}, initiatives like positionality statements~\cite{singh2025exploring} and citational justice~\cite{collective2021following, citational2022citational,citationalZine,zong2025using} have been increasingly explored, suggesting that there is space to consider the value of attribution of varied forms.
Potentially limiting or impeding the use of contribution statements is that it is unclear to what degree current systems of attribution apply to research practices in visualization and HCI. One clear example of how these systems misalign with visualization in particular is that \credit~\cite{brand2015beyond}, a popular system with strong roots in the physical sciences, has an explicit ``Visualization'' role which is intended to describe visualizations presented as figures in a manuscript, but which is particularly confusing for work in the \textit{field} of visualization.

In this work, we discuss the different formats of contribution statements that are in use as well as the benefits and drawbacks of including such statements at all.
To support our arguments, we reflect on our personal use of contribution statements, conduct  an online survey with \numParticipants{} visualization and HCI researchers in which they were asked to try attributing authorship roles to a recent paper both in their own words and using \credit, and prototype tools for surfacing and presenting author contributions.
We find and present a broad range of different opinions about and practices for contribution statements.
For instance, we find that contribution statement inclusion is particularly salient for younger scholars, scholars from underrepresented groups, and those who might otherwise experience heightened forms of bias against them.

Bringing our findings together, our recommendation is a modest, contextual one. We suggest that explicit contribution statements, written in a flexible format but guided by existing taxonomies, may be worthwhile in some contexts as they help to unravel some biases and provide credit where it is due.
More broadly, in highlighting these roles we seek to question some of the assumptions that we as a community have brought to our rendering of authorship. What work deserves credit? How should we value and evaluate contributions to a given work? There is no one answer, but we believe by surfacing this dialogue we can take steps towards a more equitable approach to doing visualization research.

\newcommand{\roleHead}[1]{\textbf{#1}. }
\newcommand{\roleSecHead}[1]{\textsc{#1}.}
\begin{table}[t]
\rowcolors{2}{white}{gray!15}
    \centering
\caption{The most popular contributor role taxonomy, CRediT~\cite{credit2026}.
    }
\label{tab:credit}
    \footnotesize
    \begin{tabular}{p{0.95\linewidth}}
        \toprule
\roleHead{CRediT Role} Definition                                                                                                                                       \\
        \midrule
        \roleHead{Conceptualization} Ideas; formulation or evolution of overarching research goals and aims.                                                                    \\
        \roleHead{Data curation} Management activities to annotate, scrub, and maintain research data for initial use and later re-use.                                         \\
        \roleHead{Formal analysis} Application of statistical, mathematical, computational, or other formal techniques to analyze or synthesize study data.                     \\
        \roleHead{Funding acquisition} Acquisition of the financial support for the project leading to this publication.                                                        \\
        \roleHead{Investigation} Conducting the research process, specifically performing experiments or data/evidence collection.                                              \\
        \roleHead{Methodology} Methodology development or design; model creation.                                                                                        \\
        \roleHead{Project administration} Management and coordination responsibility for the research activity planning and execution.                                          \\
        \roleHead{Resources} Provision of study materials, reagents, samples, instrumentation, computing resources, or other analysis tools.                                    \\
        \roleHead{Software} Programming and software development; implementation of code; testing of existing code components.                                                  \\
        \roleHead{Supervision} Oversight and leadership responsibility for the research activity, including mentorship external to the core team.                               \\
        \roleHead{Validation} Verification of the overall replication/reproducibility of results and other research outputs.                                                    \\
        \roleHead{Visualization} Preparation, creation, and/or presentation of the published work, specifically data visualization.                                             \\
        \roleHead{Writing--original draft} Preparation of the initial draft of the published work (including substantive translation).                                          \\
                \roleHead{Writing--review \& editing} Critical review, commentary, or revision of the work, including pre- or post-publication stages. \\
\bottomrule
\end{tabular}
\end{table}

\section{Background and Concepts}

Here we provide conceptual grounding for this work and connect  contribution statements with related ideas.

\parahead{Contribution Statements}
While contribution statements are uncommon in visualization and HCI, they appear frequently in publications from other fields. Authors sometimes write their statements using their own words, but are also sometimes guided by formal structures of contributorship.
Common among these is the Contributor Role Taxonomy (\credit)~\cite{brand2015beyond, credit2026}.
This taxonomy partitions tasks into 14 roles, such as Software, Methodology, Validation, and so on (see \tabref{tab:credit}).
This framing is increasingly common with some publication venues mandating its use \cite{hosseini2026enhancing}.
Allen \etal{}~\cite{allen2025contributor} found that 22.5\% of all 2024 publications with available full text in a prominent database included \credit~roles, approximately 850,000 papers.
Broadly, contribution statements seek to solve a range of different problems including bias, scientific malfeasance, and transparency~\cite{mcnutt2018transparency}---such as highlighting anti-science practices such as ghost authors~\cite{pruschak2022and}, where a researcher makes a substantial contribution and does not receive credit, or gift authorship, where authorship is given when no real contribution has been made.

\begin{figure*}[!t]
    \centering
    \includegraphics[width=\linewidth]{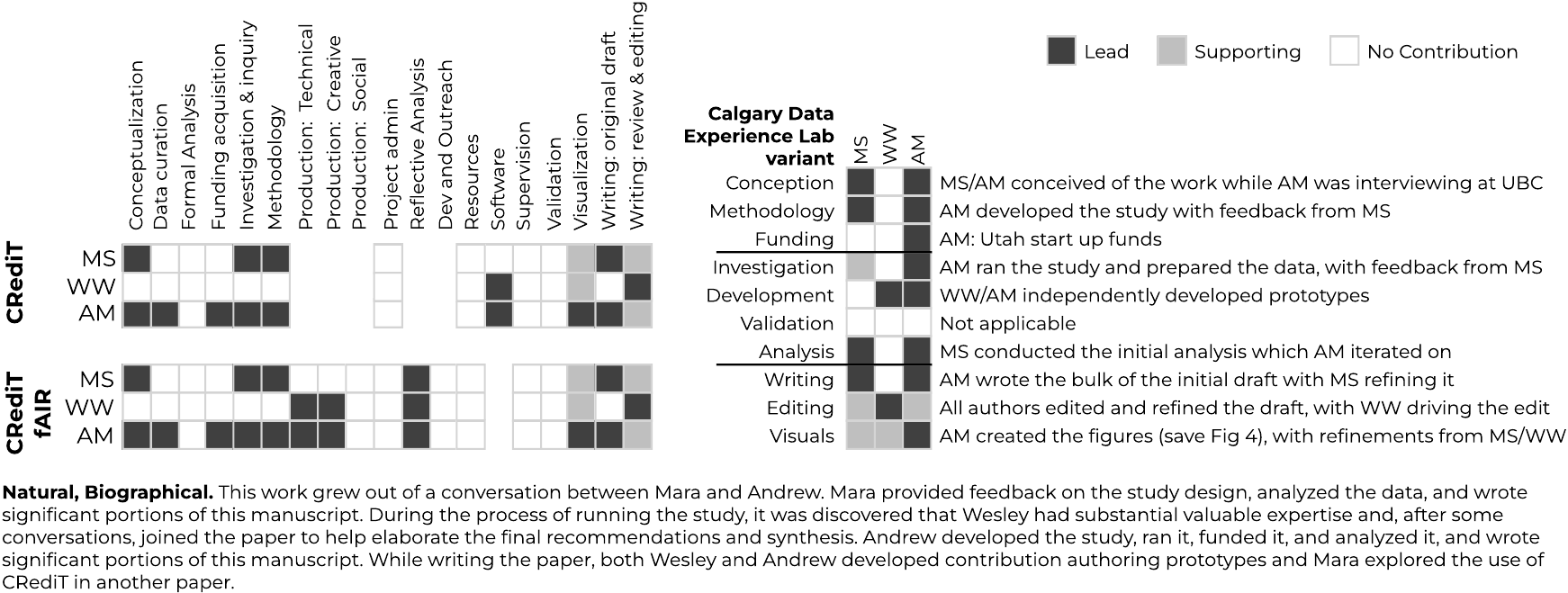}
    \caption{
    There are a range of different approaches to contribution statements, such as in these four for this work. While \credit{} and \credit{}-fAIR are typically represented textually, we show them here graphically for brevity (and to compare the differences between \credit{}-fAIR).
    }
    \label{fig:statements}
\end{figure*}

To provide a sense of what contribution statements are, we provide several different examples of them for the work culminating in this paper. While these would typically appear at the end of a paper, we show them in \figref{fig:statements} to provide concrete examples of their shape and form.
We provide a natural-language-based example, an example using \credit~\cite{brand2015beyond}, which is the most popular of the existing tools \cite{allen2025contributor} and is described in \figref{tab:credit}, as well as an alternative version to this framework for Arts Integrative Research (fAIR) called \creditfair~\cite{andrews2025beyondAuthorshipCreditin}, and a variant from the Data Experience Lab at the University of Calgary \cite{dhawka2024better, pearman2026threetopo}.

\parahead{Positionality}
Positionality statements, through which authors articulate their worldview and assumptions relevant to the work, are perhaps the most visible related example.
Singh \etal{}~\cite{singh2025exploring} explore the perspectives on these structures, highlighting the opportunity for contextualizing the motivations for the work.
Crucially, while they identify benefits of this form of statement, they also identify areas where mandated positionality statements can lead to harms of various sorts---for instance by forcing exposure of components of a marginalized individual's identity which may be stigmatized and lead to both short- and long-term damage to the individual's life and career.
We see positionality and contribution statements as being related, but not serving entirely the same purpose.
Positionality statements capture both \emph{why} the work was performed, for example to explain why four able-bodied scientists wish to study disability devices, and to \emph{contextualize} by providing relevant background details on the authors, such as the pre-existing ideas and biases of authors who are personally part of the population being studied.
These statements primarily benefit the readers by helping to clarify and justify the research. Meanwhile, contribution statements describe \emph{how} the work was performed, in particular focusing on \textit{who} did what. They primarily benefit the authors, as their contributions can be clearly and formally communicated.

To this end, we offer our own positionality statement.
The authors are engaged in this work out of a curiosity about the way the visualization community views contribution.
All have read and been authors on papers with varying numbers of authors, and have had questions about author contribution levels and whether contributions were adequately communicated via author lists and acknowledgments. \authorwesley{}, in particular, has had a longstanding personal interest in contribution frameworks and has included contribution details in most of their publications since 2023.  
Further, we see ourselves as members of the visualization and HCI communities, with two of us being in earlier career stages, and hence have a vested interest in receiving appropriate credit for career advancement.

\parahead{Qualitative Contribution Notes}
Similarly, directly describing author contributions is already expected in qualitative research paper methods sections. Qualitative researchers are often concerned with bias in their work, as methods and methodologies like thematic analysis \cite{clarke2017thematic} and grounded theory \cite{glaser1998grounded} can lead to results that are heavily influenced by the pre-existing beliefs of the data collectors and analysts.
For example, different individuals conducting un- or semi-structured interviews can have different follow-up questions, leading to different data, and different individuals coding transcripts can have different ideas about what codes and themes are important or should be grouped together, leading to different results.
Statements of varying length and detail such as ``\textit{the first two authors separately coded the first ten transcripts then met to discuss and agree on a coding scheme before the first author coded the remaining transcripts}'' are common---we make such a statement about our processes below.
These descriptions are related to positionality and contribution statements and are similarly perceived to increase transparency. Method descriptions typically seek to implicitly or explicitly convince the reader of the rigor of the research
---although the practical debiasing effects or even rigor of such approaches is unclear~\cite{barbour2001checklists}.

\parahead{Author Ordering}
Contribution is often reified in authorship and author name order, and as such it is possible to infer some contribution details from author lists---arguably this is the status quo.
However, author lists do not provide a full account of contribution, and they can be challenging to interpret due to differing norms \cite{hundley2013academic}. In visualization and HCI, it is common to use a ``sentinel'' ordering, where the first author typically contributed the most, the last author is the main supervisor, and other contributors are arranged in between, usually in order of decreasing contribution.
However, some fields use alphabetical ordering, which removes much of the contribution information from the author list.
Worse still, these approaches are applied unevenly, with some members of a given field using alphabetical ordering, while others use sentinel ordering, and still others blend the two, for example by alphabetizing all authors other than the sentinels.
Broadly, author orders have been shown to exhibit bias against underrepresented individuals---with groups like women seeing less prestigious author spots than their male counterparts~\cite{early2018understanding}. In response to this issue, feminist scholars have developed approaches and guidance for determining author lists with reduced bias~\cite{liboiron2017equity}, casting orderings as a matter of equity. Building on these ideas, Middleton \etal{}~\cite{middleton2025academic} developed the Academic Wheel of Privilege. This toolkit
helps authors externalize and visualize their privilege, reflect on this privilege,  and use those reflections to inform their work, including for author order and contribution statements.
One low-fidelity approach to credit is indicating co-first authorship~\cite{lapidow2019shared} when it is deemed that two authors did equal work.
While this superficially addresses one frequent source of author-order tension, there are often still political concerns about which author's name \textit{actually} appears first.
A somewhat rarer pattern is indicating co-senior authorship, in which the last n authors are indicated as having done equal work~\cite{gebreegziabher2025supporting}.
These frictions suggest that there is an underexplored landscape of potential annotative caveats, mediated by daggers and asterisks, waiting to be unpacked.

We spent significant time considering and discussing what the most appropriate author order would be given the above distribution of the work. We ultimately selected a ``sentinel'' order that placed \authormara{} and \authorandrew{} in what are traditionally the two highest-prestige positions in visualization papers, and assigned \authorandrew{} the last author to reflect the PI-relationship he had with the study.

\section{Process}

We used three methods to consider contribution.

\parahead{Reflection} First, we reflected on our uses of contribution statements, which all three authors have used to varying extents and in varying formats in the past. We also reflected on contribution statements for submissions we made while writing this paper, and had asynchronous discussions about our thoughts and decisions.

\parahead{Survey} Second, we conducted an online survey of the academic visualization community on the role of contributorship.
After consent, the survey asked participants to identify the title and authorship of a recently submitted visualization or HCI paper that they were a co-author on.
The survey asked them to list all of the tasks and roles that each author took on that paper in their own words.
Next, the survey showed the \credit~taxonomy and asked participants to again list roles that each author took on, but using that framework.
Finally, it asked participants to reflect on the experience of assigning authorship, as well as their views on contributorship and related topics, such as whether or not the \credit~taxonomy was a good fit for visualization work.
The survey allowed participants to repeat the identification and assignment steps as many times as they wished---although only one participant did the exercise more than once.

As this study is meant to be informal, it was conducted within a Word document (which allowed participants to revise earlier answers, a design limitation), which was conveyed to participants via direct message or email.
We identified participants through a convenience sample of \authorandrew's recent collaborators,  which was then extended via snowball sampling.
This study was marked exempt by the University of Utah's IRB (\#00203562).
We provide an unmodified version of the survey in the supplemental materials, available on OSF: \href{https://osf.io/xyk9p}{osf.io/xyk9p}.

Through this method, we recruited \numParticipants{} participants, out of 25 participants directly solicited and posted in two small Slack channels.
Participants had a range of academic career stages---Ph.D. student (6), postdoc (3), assistant professor (5), associate professor (5), and full professor (2)---and used a range of pronouns---he/him (10), she/her (7), they/them (2), he/they (2).
In compensation for their participation, for each participant we donated 10\$ USD  to the Internet Archive.
Participants were qualified to engage with the study if they had submitted a visualization or HCI paper in the last six months.
For privacy reasons, the data can not be shared. Authors, who are all part of the same community, may have shared sensitive information and opinions and may hold diverging views on who did what, and so we do not wish to risk conflict or perspective exposure.
Accordingly, participant quotes
are unlabeled to increase anonymity, which some participants worried greatly about.

\begin{figure}[t]
    \centering
    \includegraphics[width=\linewidth]{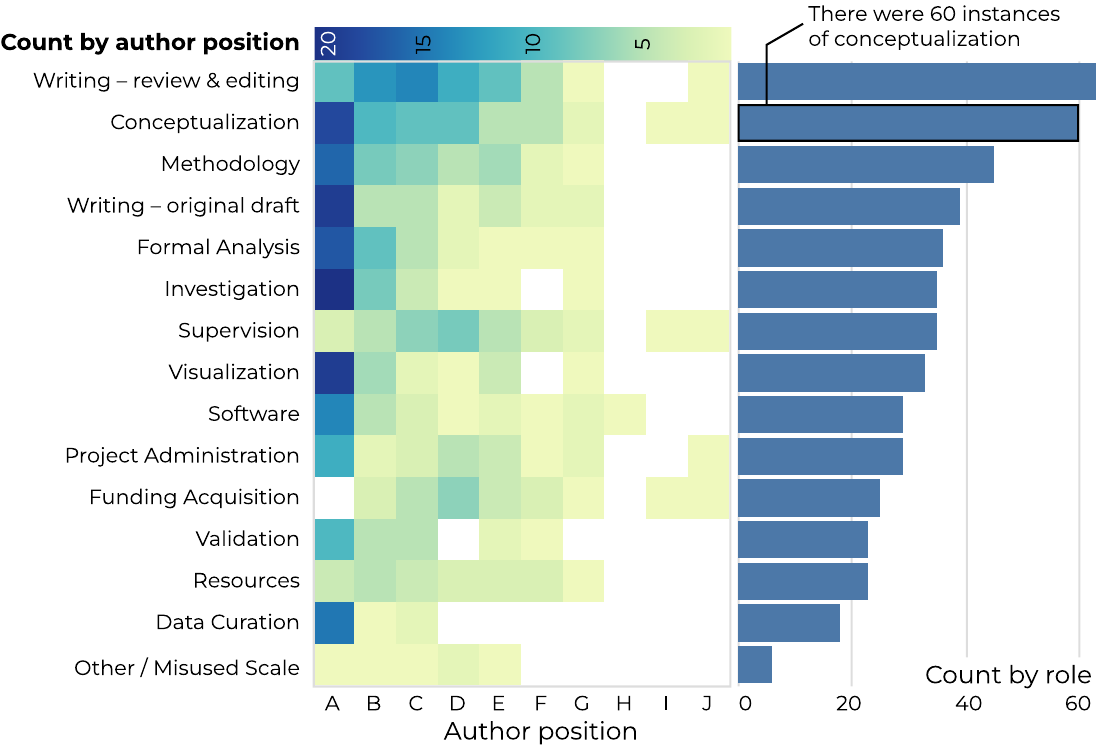}
    \caption{Credit attribution by category for participants in our study. While these results point to editing and conceptualization being the most commonly noted categories,
we stress that these findings are merely gestural and not conclusive.
        Moreover, we also note that summative counts of this nature tend to be reductive (for instance the lead/supporter/none ranking from \figref{fig:statements} would not fit into this summation).
        We include this as a means to characterize our instrument, rather than to make any strong claims about contribution patterns.
}
    \label{fig:credit-cat}
\end{figure}

\authormara{} and \authorandrew{} separately analyzed participant data and then asynchronously discussed the results.
Given the recruitment techniques, any quantitative analysis would not be representative of the entire community, although we provide a gestural summary in \figref{fig:credit-cat}. Instead, we focus on qualitative analysis, characterizing the breadth of opinions and practices espoused. \authorwesley{} was initially a study participant, but joined the author team after the analysis was completed.
In the interest of transparency, we quote \authorwesley{}'s survey responses \wwqt{like so} and other participants \qt{like this}.

\parahead{Prototyping} Finally, \authorandrew{} and \authorwesley{} independently developed lightweight digital tools for scaffolding the process of describing and presenting contributions. The authors all discussed these tools, noting their similarities and differences, and used this discussion to promote further thinking about our earlier reflections and survey results.
\authorandrew's tool, shown in \figref{fig:toolsnapshot}, showcases a more representation-neutral approach, supporting graphical input of contributions using several existing taxonomies and producing contribution lists and tables in a number of formats.
This design draws significantly on Holcombe \etals{}~\cite{holcombe2020documenting} tenzing, which supports development of contribution lists, as well as Middleton \etals{}~\cite{middleton2025academic} Academic Wheel of Privilege.
\authorwesley's tool, which is shown in \figref{fig:westoolsnapshot}, 
extends the contribution matrix format used by the Calgary \credit{} variant and uses it as an authoring framework for creating tables with open-ended taxonomies and contribution levels.
Public versions of both tools are available at \href{https://contrib-builder.netlify.app/}{contrib-builder.netlify.app} and \href{https://wjwillett.net/misc/matrix-builder.html}{wjwillett.net/misc/matrix-builder.html}.
Both tools produce LaTeX-compatible output, and we hope they can serve as a starting point for other researchers who wish to surface author contributions into their own work---or for venues that wish to encourage contribution statements.
They also highlight the potential for further interactive authoring and visualization tools that might help researchers collect, reflect, and share contributor information in more productive ways.
Both tools made significant use of AI coding.

\section{Structuring Contribution Statements}

We discuss the different formats and visual representations of contribution statements that are in use, then describe our own contributions to this paper in various ways.

\subsection{The Structure of Credit}

\credit~is popular, and its use is required in certain publication venues---yet it often feels misaligned with the broader epistemologies and norms of visualization and HCI-oriented research.
Several participants noted that having a taxonomy was useful for attribution, as it helped remind them of the work that different contributors took on.
However, others found \credit{} difficult to use or poorly fitting for visualization and HCI.
Some mentioned that \credit~was too detailed, and that it took significant time and effort to properly use the framework. Others described the framework as too broad, and stated that they wanted finer-grained roles.
Wesley suggested that it drew from a fundamentally different disciplinary origin, observing \wwqt{I find \credit~just smacks of a bunch of assumptions that are anchored in the physical sciences.}
For instance, multiple participants pointed to the role titled \textit{Visualization} as one of the culprits of the framework's poor fit, as the role, which was originally intended to describe visualizations used in paper figures, is quite confusing when used in the context of the field of visualization. Some participants noted that this role could simply have a more precise name, such as ``paper figures'' or ``visuals''.

Many others mentioned that there were a range of roles that are common in visualization and HCI that were not captured by \credit. For instance, while conducting experiments and collecting data is listed as a role in the framework, the design of such studies, applications for ethical approval, and participant recruitment are not clearly indicated.
In the fields of visualization and HCI, the creation of a design to study or as a result of study findings is common. However, the roles of design and design feedback are also absent from the \credit~taxonomy, forcing many of our participants to report these tasks under roles such as \textit{software}, which is focused on implementation, or \textit{resources}, which is focused on providing study materials.
Another common aspect of design work is collaboration with domain experts, who often become authors on the papers as well, but domain expertise is challenging to categorize using the framework, such as in the case of \qt{co-design where there are specifically different expertises brought to the same activity}. To that end, another participant underscored that participant recruitment is difficult \qt{because working closely with communities is time consuming and a contribution in itself that requires time, dedication and skill}, which is not well reflected in the \credit~approach.
Similarly, another participant suggested that \qt{I also don't think it captures conceptual/theoretical contributions, where the contribution comes from more reflection and framing than data analysis.}
Finally, some participants mentioned a desire for a separation of different types of work and contribution, for example by making clear divisions between quantitative and qualitative investigation and differentiating analysis of collected data from other types of research tasks like reflection. For instance, one participant mused that under \credit~\qt{Qualitative analysis is definitely ``formal analysis'' in that you are applying a known methodological process to produce conclusions even though it's ``other''}.

Contribution statements can take many different forms, and \credit~is not the only existing standard.
For instance, the \creditfair~\cite{andrews2025beyondAuthorshipCreditin} taxonomy (\href{https://groundworks.io/credit-fair}{groundworks.io/credit-fair}) is a variation of \credit~specifically made for arts and arts-integrated research \cite{andrews2025beyondAuthorshipCreditin}. While \creditfair~addresses some of the issues that our participants mentioned, including adding explicit creative roles that could describe design work, one of the two participants who mentioned having used it said they still found it hard and unsatisfying to use.
\authorwesley{} even responded in the survey that \wwqt{In practice, I'm not convinced that it's really possible to find a single systematic taxonomy that will be work broadly even within our field.}
Free-form contribution statements
are another alternative, and naturally permit more freedom in how authors describe their contributions---yet this flexibility also makes them more difficult both to parse and construct. For example, as noted above, frameworks are helpful for reminding authors of all of the forms of work that co-authors might have performed.
Examples like those from Wesley's Data Experience Lab at Calgary (see \figref{fig:statements}) blend several of these approaches---tuning the \credit~types to better align with visualization and HCI work, introducing multiple contribution levels, and incorporating free-form notes (\figref{fig:westoolsnapshot}) to provide additional detail and context.



\subsection{Visual and Interactive Approaches}

As fields that deeply consider the presentation of information and its social implications, the visualization and HCI communities are uniquely positioned to innovate how contribution statements are handled---creating new ways of visually communicating contributions and new tools for articulating them.
Publications using the Calgary \credit{} variant~\cite{dhawka2024better, pearman2026threetopo}
have presented contributions as matrices (inspired by examples from the neuroscience community~\cite{tay2021researchers})
to provide more glanceable assessments of contributions.
Related graphical forms such as a biofabric~\cite{longabaugh2012combing} or text visualizations~\cite{brath2020visualizing} might make these structures easier to parse.
Akbaba \etal{}~\cite{akbaba2026designing} include a bespoke contribution statement that interweaves a phased timeline with contribution roles, yielding a situated credits page for the work. 
More generally, we echo Varona \etal{}~\cite{varona2025theory} in that bringing varied visual forms to the representation of contributions can shape how we conceptualize contributions. Accordingly, we underscore the value of continuing to explore this design space.
Further, interactive tools that support the creation, evolution, and sharing of contribution statements have the potential to make these practices accessible to groups and communities that might not previously have considered them.

\section{Should we require contribution statements?}

Participants had varied views on contribution statements, from agreement (sometimes with caveats), to uncertainty, to refusal.

\subsection{Pro-Inclusion} Participants saw a range of benefits for including contribution statements, which varied based on their context.

\parahead{Increasing transparency}
Transparency to readers, who may include potential future employers or collaborators, was an ongoing theme. For instance, \qt{this is another aspect of transparency that we could be in the lead for aiming to do, no different than trying to publish reviews, disclose AI use, [write] reproducibility statements}.
Another respondent emphasized that including contribution statements would be valuable \qt{because the discipline has unstated inconsistent and unfair ways of allocating authors and exposing or hiding contributions and anti-contributions.}
This aligns with Wislar \etals{}~\cite{wislar2011honorary} observation of a high amount of gift and ghost authorships in biomedicine. McNutt \etal{}~\cite{mcnutt2018transparency}, for instance, suggest that authorship attribution is one antidote to these issues.
Aligned with this perspective, one participant suggested that including these statements would \qt{reduce the author bloating}---which in their case was consistent with gift authorships.

One area that explicit contribution attribution opens is greater and more nuanced dialogue around author order.
For instance,
one participant noted that one time they \qt{wrote down [their] contributions in order to argue for [their] first-author position}.
By specifically explicating what work was done, authors have the opportunity to identify tasks that were forgotten or misinterpreted, or simply to provide concrete evidence to make a case for a particular authorship.
These findings align with previous work showing how \credit~can help to resolve authorship disputes \cite{partin2025using}, particularly between junior and senior authors.
However, this is not a panacea.
One participant offered a realistic commentary to this effect, observing that \qt{Credit is a ``sensitive'' topic, and ultimately, advisors have the final say. So if the advisor disagrees with how the main contributor lists out a contribution via the \credit~taxonomy, I think the same underlying issue will persist. }
Anywhere there are scarce resources, which in our case are the \emph{prestigious} sentinel authorship positions, there will naturally be tension in the allocation of those resources, which is often mediated by power dynamics, as here.
One value of the transparency offered by contribution statements, then, is the potential to renegotiate those dynamics through externalized evidence.

Crucially, however, these statements are self-reported and there is no mechanism to prevent omission, fabrication, or authorship misconduct. 
This factor means that there is a limit to the value of transparency mediated through contribution statements. As one reviewer of this paper summarized, ``we can only trust these statements to the same extent we trust the people who wrote them'', suggesting that their most poignant value might be elsewhere.


\parahead{Providing due credit} One group that was particularly enthusiastic about contribution statements were those with less structural power or positional stability, which often manifested as
those in earlier stages of their career---echoing
Liboiron \etals{}~\cite{liboiron2017equity} observation that ``women and junior researchers [...] consistently receive less credit for equal work''.
For instance, one Ph.D. student respondent stressed that
\qt{I believe [Contribution Statements] should be added as a mandatory requirement. A lot of the time while writing papers, there are factors that are external to the amount of contributions or roles of each co-author, which can affect the ordering of author list in the paper}.
Similarly, an early assistant professor observed that contribution statements \qt{could shine light on exploitative practices between advisors and students, where students do most of the work but are not getting recognized as ``independent researchers'' because their supervisor is still listed as a co-author.}
We suggest that those who are more precariously placed gain greater benefit from clear contribution statements.

One participant said:
\qt{I'm thinking of doing this for all the papers I lead from now. I really like having this: claim my credits (ouch for my career), highlight my student contribution (for them, they are not just executing ideas, they have ideas and take the leadership).}
indicating that such statements can help to address power imbalances between students and supervisors in the author team~\cite{akbaba2023troubling}.

\parahead{Promoting reflection} One potential value of contribution statements may be merely in assembling them, regardless of whether or not they end up being included in the final manuscript.
For instance, one participant said that writing a contribution statement can \qt{help students understand the authorship}.
Similarly, Wesley described contribution statements as being useful primarily as \wwqt{an evolving tool that serves internal pedagogical, methodological, and accountability ends within research teams and provide a mechanism for communicating and acknowledging individual authors' roles in the work.}
Other participants discussed how explicit role discussion during a research project can improve the work and the relationships between authors. Beyond just understanding, communicating, and acknowledging authorship, explicit creation of contribution statements may facilitate the negotiation of authorship and author order.
This process-centered perspective highlights the value of having a scaffolded means to reflect and consider authorship and contributorship. 
Recently, Parsons and Shukla~\cite{parsons2025beyond} discussed the benefits of reflection in the design process, and Varona \etal{}~\cite{varona2025theory} discussed how frameworks and other shaping tools can support reflection. We believe this recent emphasis on reflection also applies to the idea of contribution and authorship.

\credit~and its variants are sometimes suggested as a means to surveil or account for work. However, \authorwesley{} noted that \wwqt{I'm much less convinced by their utility as tools for systematic external accounting or analysis.}
Inevitably, any enumeration of work, such as counting first author papers, may be used in this fashion.
The enumeration may lead to a perspective in which whoever simply did the most roles is seen as the primary author of the paper, potentially devaluing the work of others.
One participant noted that \qt{In those cases, having a \credit~taxonomy say only `conceptual' would seem unfair (as in the end in \credit~you tend to count/see who does most tasks rather than count significance/uniqueness of those tasks).}
These sentiments speak to the potential benefit of more open-ended systems for documenting contributions---which can make it easier to describe novel approaches and reconcile differing roles, but comparisons across papers are more difficult.
Echoing DORA's~\cite{american2012san} call for evaluating research quality qualitatively rather than quantitatively, we highlight the value of contribution statements as qualitative artifacts rather than means to extract discriminative numerics.
In any case, we stress that research teams may find value in the process of articulating contributions as a way of collectively negotiating authorship, even if they forgo publishing them.

\begin{figure}[t]
    \centering
    \includegraphics[width=\linewidth]{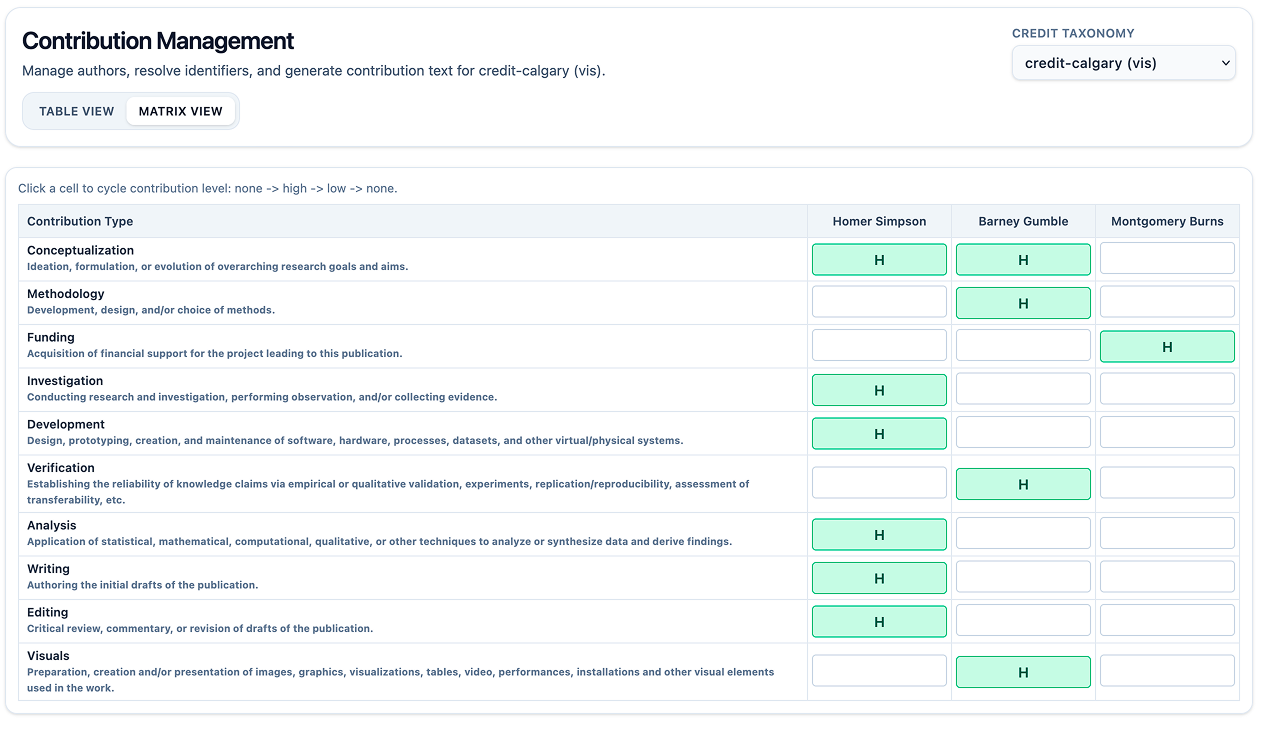}
    \caption{\authorandrew{}'s contribution and authorship solicitation and ordering prototype, available at \href{https://contrib-builder.netlify.app/}{contrib-builder.netlify.app}.}
    \label{fig:toolsnapshot}
\end{figure}

\subsection{Anti-Inclusion}

On the other side of the coin, some respondents highlighted issues or hesitations with including contribution statements.

\parahead{Failing to add value}
Some participants suggested that it was unclear if contributions would address any essential issues. For instance, one noted that \qt{I don't think it would hurt, but also I don't think it's a solution to a problem.}
Another participant speculatively questioned what problems these statements might solve:
\qt{Is the goal to get people to think about authorship intentionally before submission? Is the goal to prevent bad actors hanging their names on stuff they didn't do? Is the goal to increase clarity and decrease weird edge cases where people are unhappy?} 
Indeed, one perspective~\cite{mcnutt2018transparency} frames contributions as a tool to address issues like ghost and gift authorship by increasing transparency, although the relative effects on other issues are less clear.
Another participant noted that \qt{I would not mind it but to be completely transparent I have never felt the need for them. But... if presented with an argument for I might support it.} This statement seems to suggest resistance may be a matter of awareness rather than intended convention, a distribution that this paper seeks to realign.

A related sentiment is that some participants believed that contribution statements do not address bias as well as sometimes claimed. While readers of papers without contribution statements may be biased towards seeing certain authors, such as those in the first or last position, as the most important, contribution statements may just redirect that bias towards specific roles which may not be viewed equally. \qt{I could see it causing issues and bias where some types of contributions (e.g. hard skills/soft skills) are seen as more/less valuable than others, and therefore some people are seen as better/worse researchers/contributors. That can have downstream effects on tenure, hirability, etc. ALL types of contribution are necessary and important.}
This echoes Bowker and Star's~\cite{bowker2000sorting} observation that ``like any classification scheme that renders work visible, it can also render surveillance easier''.
One participant argued further that, if the contribution of authors is under question, then contribution statements may not help, as a reader could dissect and interpret the statement in a way that aligns with their current thinking.

\parahead{Shifting the problems}
Another potential issue is that contribution statements may not be written accurately. As we observed in the survey responses and our own use of contribution statements, it can be easy to forget that an individual held a role or that a role existed at all when trying to recall events at the end of a project---leading to unintentionally inaccurate contribution statements. There are also situations where social pressure leads to \textit{intentionally} inaccurate contribution statements. For example, authors may find it awkward to show that a co-author held fewer roles than other co-authors. Junior authors (like students) may feel pressure to attribute more roles to senior authors or supervisors
to maintain a positive relationship.

The process of writing contribution statements can cause issues as well. Given the breadth of opinions on the value of contribution statements, bringing up the topic of including them could reveal disagreements amongst the author team, which could lead to friction.
Even if all authors agree to include a contribution statement, explicit discussion of roles can surface different interpretations of the project, which can cause friction and tension between authors.
These tense power dynamics can lead to some unexpected outcomes, as one participant recalled:
\qt{My first important paper was single authored because I didn’t want to burden my colleagues (supervisors) with my dumb ideas and was too scared to talk to them. They didn’t talk to me about authorship and credit ... so I published solo. I regret this}.
Scaffolded community norms like explicit authorship discussions may help circumvent this form, or others, of regretted outcome.

\parahead{Adding chores}
Finally, some survey responses described contribution statements as a chore, stating that adding yet another piece of paperwork to the paper submission process is annoying.
One participant characterized filling out contribution statements as \qt{It's a chore, and academics have enough self-assigned chores.}
This echoes the growing amount of non-research work that is required of authors in the paper publication process, including interfacing with processing systems like ACM TAPS~\cite{taps} which ostensibly provides a path for making the rendered documents accessible, writing impact or ethics statements which support ``\textit{anticipation}, \textit{action}, \textit{reflection and awareness}, and
\textit{coordination}''~\cite{prunkl2021institutionalizing}, requiring DOIs in bibliographies~\cite{visDois} which aim to ensure that tracing references is straightforward, \etc{}
While there are always reasons for each additional requirement, their sum  can be daunting and frustrating.

One participant observed that \qt{In my opinion these seem like minor wins at best for a relatively minor chore.}
Colloquially, in discussing this work with senior PIs outside of visualization, some noted that they had never read or paid attention to these statements, arguing that their value was limited---we suggest that this may in part be because there have not been sufficient incentives for those with power to engage with them and appreciate what they can reveal.
Some participants felt that contribution statements \textit{were} valuable, but that their value may not be worth the effort. One participant related this issue to how effective and efficient the contribution statement expectations were, saying that if the \qt{mechanism is not carefully designed, it has the danger to become a trivial administrative procedure that doesn't have much value.}
Balancing the use of contribution with appropriate incentives, or lack of disincentives, is critical to realizing the potential value of these statements without causing them to be dismissed as tedium.

\section{Recommendations}
Informed by these reflections, \textbf{we recommend that visualization and HCI venues should encourage contribution statements, exempt them from page limits, and provide flexible templates, examples, and justifications that support their inclusion}---but we should not require them or force the use of any particular format.
For venues with rigid word or page limits, we suggest that statements be permitted to spill onto the reference pages. similarly to how ethics statements and acknowledgments are currently handled in IEEE VIS papers~\cite{visDois}---allowing them to appear in the main document rather than as supplemental materials.
Echoing considerations from fields that have required AI ethics or impact statements such as in NeurIPS, there are substantial opportunity costs~\cite{Ashurst_2022} to requiring these structures when not exempted from that limit, as ethics and impact statements were at NeurIPS in 2020.
Allowing them to be optional without making space for them creates a usage disincentive.

Crucially, we do not suggest that these statements be required, but be encouraged---especially in scenarios where they will have the most value. This recommendation aligns with how positionality statements are currently treated in visualization and HCI.
For instance, solo author papers would have little use of such statements---a position echoed by one participant: \qt{I wonder if \credit~fits all types of paper, like a review paper? Or when there is just 1 author?}
We stress that institutional encouragement, such as by providing examples in a call for participation or by them not counting towards page lengths, normalizes inclusion of such statements and does not rely on authors opting to spend space on them.

To that end, we suggest venues avoid prescribing a specific format for contribution statements, instead providing examples of a variety of established models and acknowledging the validity of free-form statements and bespoke taxonomies.
We further suggest that visualization and HCI venues should describe these options on their websites---promoting the idea of contribution statements and emphasizing their value not just as an accountability mechanism but also as a helpful tool for navigating discussions about author order and value.
There \emph{may} be a danger in the systematic use of contribution statements as it may lead to unintended quantified uses of author roles, echoing the issues with h-indices~\cite{american2012san}---however, we leave deeper consideration of these dangers to future work.

This non-prescription opens a range of pressing questions: how granular should the delineated roles be? And how should specifics be reconciled with standard role taxonomies? Standards make work legible but they can also erase things not understood by the standard.
We underscore that it is unlikely that a single taxonomy will work for all cases, or even for all of visualization---experiment analyses have fundamentally different project structures than design studies and exhaustively trying to attribute all roles to all authors may be uninformative.
We suggest that the most effective form will be aligned with the audience and context. Such selections might then be either tweaked or annotated (again like those of the Data Experience Lab) to clarify or augment specific roles.

Lastly, we suggest that our community continue to consider and reflect on its assumptions, as has been ongoing recently \cite{saharan2026critical}, including in any approach that is taken towards contribution statements. Just as Ashurt \etal{}~\cite{Ashurst_2022} highlight with respect to AI ethics statements, there remains a need for good examples and best practices for authoring and sharing contribution statements effectively.
To that end, one participant explicitly called for more community engagement---\qt{It would be a great thing to workshop – a really important thing for the community}---a position we echo. We see this work as an early comment in a longer dialogue.

\section{Limitations}
The informal nature of the survey, in which participants were not disallowed from altering earlier answers as they worked through the study, may have affected responses. For instance, in piloting it was noted that the \credit~component reminded some participants of roles that some authors had performed, which inspired them to change their prior responses.
The initial participant sample was also drawn from Andrew's recent co-authors, and those connections may have tilted responses.
More generally, participants willing to engage in this study may have been more positively biased towards it---for instance, one researcher  explicitly declined to participate, objecting to the idea of contribution statements.
Lastly, this work makes a call to action, which is that we should consider contribution statements in visualization. The notion of contribution is studied widely elsewhere, and this work seeks merely to provoke a discussion within this community.

\section{Discussion}

Here we give consideration to a series of concerns and areas for future work highlighted by this work.

\parahead{Memory, allocation, and mediation}
One issue we observed was that some participants struggled with effectively characterizing who did what on a paper.
For instance, for a paper on which one of the authors of this paper was a co-author, there were a number of identified gaps, misrecollections, and misapplications of the \credit~taxonomy.
While this was likely related to the artificial nature of this study, as preparation of a contribution in a paper would likely receive more care and attention, it highlights a potential danger: people may forget or misattribute work.
Beyond the human propensity to make mistakes, this challenge might become especially salient during long-running projects that take place over years.
Further, another participant noted that there may be a tendency to seek complete assignment of all roles, when not every project has all roles in it---for instance, a theory paper may not involve any software at all. Similarly, authors may be tempted to treat allocation of contributions as a quantitative exercise, or worse, a zero-sum one, leading to contribution statements that systematically misrepresent contributions.
Some participants (including \authorwesley) highlighted how many of these issues might be addressed by treating the authoring of contribution statements and selection of author order decisions as collaborative discussions within the research team. These conversations, mediated through a concrete contribution taxonomy or artifact, can provide opportunities for learning, reconciliation, and recognition among collaborators that surface assumptions about value which are often unclear---especially to junior researchers.
While group care and focus offer checks for these types of issues, there will inevitably be mistakes and missteps just as there are with any other part of human-produced documents. Accepting that these may happen and potentially providing mechanisms for post-hoc correction seem like appropriate salves.

\begin{figure}[t]
    \centering
    \includegraphics[width=\linewidth]{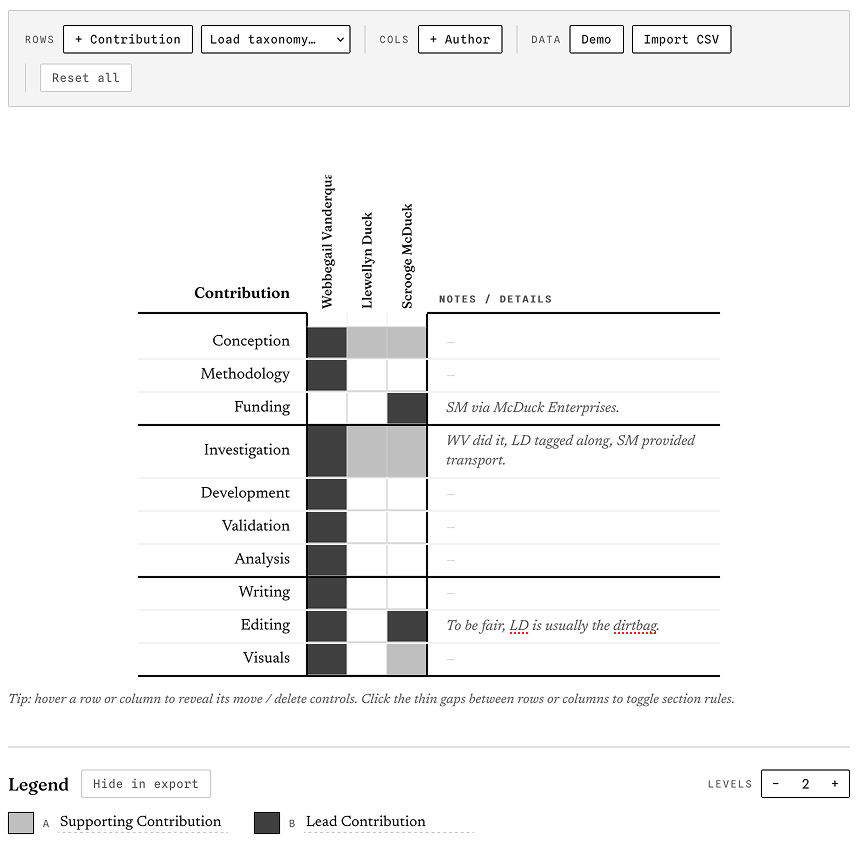}
    \vspace{-1em}
\caption{\authorwesley{}'s contribution and authorship solicitation and ordering prototype, available at \href{https://wjwillett.net/misc/matrix-builder.html}{wjwillett.net/misc/matrix-builder.html}.}
    \label{fig:westoolsnapshot}
    \vspace{-1em}
\end{figure}

\parahead{Exploring the design space of contribution types}
While taxonomies like \credit{} and \creditfair{}  offer structured formats for documenting contributions, their fixed structure and binary (\textit{did}/\textit{did not}) designations offer a particularly narrow and coarse-grained view. As a result, they are unlikely to capture the specific contribution types for any given field or project well, and may still flatten or obscure contributions from individual authors.
Addressing these concerns likely calls for the creation of new structured and unstructured frameworks for characterizing contributions.
For instance, the Calgary \credit{} variant  drops or modifies a number of categories to better align with visualization and HCI research, and also introduces an ordinal distinction between ``lead'' and ``supporting'' contributions. Yet the space of possible contribution types, levels, and so on is considerably broader, and would benefit from systematic exploration of taxonomies, visual representations, and authoring approaches---all issues our prototypes (Figures \ref{fig:toolsnapshot} and \ref{fig:westoolsnapshot}) gesture towards.

\parahead{Criteria for authorship}
Perhaps the most consequential dividing line when considering contributions is whether an individual is counted as an author at all. To that end, we heard several different perspectives on what counts as authorship, which in turn invite differing perspectives on the necessity of contribution
statements. One common framing offered authorship to \qt{whoever has intellectual contribution}, which we also heard framed as \qt{would this paper have happened in this form without this person?}. We note that this involves an arbitration of power, often mediated by a PI, through which gift authorships might also be bestowed---such as to the negligent secondary advisors that one respondent bemoaned.

Participants also highlighted diverging notions of what is considered attributable labor.
For instance, \qt{I consider UGs who do labeling data, writing code, and doing low-level jobs qualified for authorship if their contribution is substantial (\eg{} we actually used their output, they did more than 10 hours). My [PhD advisor] disregards pure low-level contribution, but I think high-level contributions come from low-level work.}
Just as with author order often being the decision of those in power, so too is authorship. A closely related model is the fixed work/fixed pay model, in which potential authors must participate for a fixed number of hours, perhaps 10 or 20.
Another participant espoused differing criteria, which they referred to as the ``Hollywood'' model: \qt{I think anyone who contributes should be an author. My background is in the film industry, where the movie credits are endless and everyone gets a credit no matter their role... It costs nothing to give credit, and it can mean the world to be recognized and to get credit.}

While not able to directly reshape the power dynamics or to fundamentally alter perspectives on the value of labor, having the thoughtful conversations around attribution that contribution statements require may be enlightening to those involved.

Cases on the border of authorship are particularly challenging. 
Participatory research often struggles with the question of whether participants and other non-academics should be authors on resulting publications \cite{sarna2017where}. 
On multiple occasions, one of the authors of this paper (Andrew) has had potential study participants request authorship as a prerequisite for participation. While they were turned down, this highlights the limitations of attribution via authorship. 
Is a one-hour engagement worth authorship? What about a lifetime of experience consulted and extractively reused? 
The answers to these questions vary, but we suggest that the default perspective on authorship and contribution limits the nuance of discussion of such variance. Further complicating this potential means of compensation is that authorship necessitates identification, which conflicts with anonymization norms.
However, Zong~\cite{zong2025using} argues that participants need not always be anonymized, characterizing the direct naming of study participants---in cases when such involvement does not warrant authorship, particularly for disabled study participants---as a form of citational justice. However, naming study participants carries risks as well, as some may not want their actions, words, and beliefs at a specific point in time publicly tied to them forever.
We suggest there can not be one universal answer to these questions (Should I name my participants? Should a source count as an author?) because they must be answered contextually and in dialogue during the work and with the community that work occurs within. 

\parahead{Role of AI} Early in the discussions that lead to this paper, we brought up the impact of AI and how its use could be represented in contribution statements, but we did not come to conclusive results. One participant accurately described the current milieu by provocatively wondering, \qt{AI is increasingly used to contribute to many of these activities named in \credit. Does that change anything? In one view, an author is making that contribution using a tool. In another view, the AI-enabled author is not doing the same epistemic work as before and is also using a system that is epistemically indebted to all of the content on the internet.}
They highlight that some works would not have been completed without this form of assistance, which echoes the relatively common authorship criteria to that effect, suggesting that there may be space to consider AI tools as having meaningfully contributed to a work.
This reformulation of the agency of particular tools may usefully provide new means through which to consider human authorial intention and agency. More pragmatically, it may also blur the lines between contribution statements and AI use or ethics statements, necessitating compatible approaches for acknowledging both kinds of research inputs.

\parahead{Relationship with anonymity}
A long-running debate is whether there is value to anonymization of authorship during submission.
It is easy to be biased by a certain name appearing in an author list and to give affordances to that work that might not be granted otherwise. For instance, it's possible someone might suggest \emph{``I know Moe Sizlack makes great systems, I don't necessarily have to check the code as deeply''}.
Complicating this debate is that in some venues, such as IEEE VIS, most of the reviewers \emph{do} know the identities of the authors as two of the minimum three reviews are completed by PC members, potentially limiting the utility of the anonymity process, if authors elected to be anonymous at all.
Explicit contributor lists may help to unwind some of those biases by making it clear what the reader should expect from their understanding of the authors involved.
For instance, in the aforementioned paper involving the potentially great system, Moe might have only provided some resources, such as meeting space.
However, the relationship between anonymity, contribution statements, and bias is unclear and warrants further study.

\section{Conclusion}
This work explores the visualization community's perspectives on contribution statements via reflection, a survey with a small slice of visualization and HCI researchers, and prototype design.
We do not claim to speak for the whole community, nor that our sample is complete or representative.
Instead, we offer the position that contribution statements are an interesting and useful construct that authors should consider both during the research process and while writing papers.

There is no one perfect solution for addressing the issues of transparency, bias, and scientific malfeasance. Contribution statements offer an explicit approach to these issues, but there are few guardrails to prevent contribution falsification or, perhaps more likely, simply inaccurate memory and reporting.

A growing thread~\cite{akbaba2023troubling, zong2025using} of research pushes us to care for our participants and collaborators.
We echo that perspective, and emphasize that we should apply a similar lens of care to our co-authors.
We might manifest such care, as a community, by endeavoring to be thoughtful in our approaches to attribution and credit.

\acknowledgments{
We thank our participants for their time and consideration.
We also appreciate pointers and commentary from Derya Akbaba, the Utah HAVOC lab for piloting the study, feedback from the UBC InfoVis lab, and our reviewers for their thoughtful commentary.
We also thank past and present members of the Data Experience Lab and other collaborators (particularly Jason Dykes) who helped evolve the Calgary CRediT variant.
This work was supported in part by the Canada Research Chairs program.
}

\bibliographystyle{abbrv-doi}

\bibliography{template}



\clearpage

\appendix


\end{document}